\documentclass{elsart}
\usepackage{graphics}
\begin{document}
%**************************************************
%EQUATIONS & ENVIROMENTS
%************************************
\def\re#1{(\ref{#1})}
\def\beq{\begin{equation}}
\def\eeq{\end{equation}}
\def\beeq{\begin{eqnarray}}
\def\beeqn{\begin{eqnarray*}}
\def\eeeq{\end{eqnarray}}
\def\eeeqn{\end{eqnarray*}}
\def\nome#1{{\label{#1}}}
\def\ln{large-$N\;$}
%*********************************************
%**********************************************
%GREEK LETTERS
%********************************************
\def\a{\alpha}
\def\b{\beta}
\def\g{\gamma}                  \def\G{\Gamma}
\def\de{\delta}                 \def\D{\Delta}
\def\e{\varepsilon}
\def\k{\kappa}
\def\l{\lambda}                 \def\L{\Lambda}
\def\m{\mu}
\def\n{\nu}
\def\ta{\tau}
\def\r{\rho}
\def\s{\sigma}                  \def\S{\Sigma}
\def\th{\theta}                 
\def\x{\xi}                     \def\X{\Xi}
\def\y{\upsilon}              \def\Y{\Upsilon}
\def\z{\zeta} 
%********************************************************************
%CALLIGRAPHY
%%%%%%%%%%%%%%%%%%%%%%%%%%%%%%%%%%%%%%%%
\renewcommand{\AA}{\mathcal{A}}
\newcommand{\DD}{\mathcal{D}}
\renewcommand{\SS}{\mathcal{S}}
\newcommand{\GG}{\mathcal{G}}
\newcommand{\NN}{\mathcal{N}}
\newcommand{\OO}{\mathcal{O}}
\newcommand{\PP}{\mathcal{P}}
\newcommand{\ZZ}{\mathcal{Z}}
\newcommand{\WW}{\mathcal{W}}
\newcommand{\zk}{\mathcal{Z}_{k}}
\newcommand{\wk}{\mathcal{W}_{k}}
\newcommand{\wka}{\mathcal{W}_{k,a}}
%*********************************************************************
%SPECIAL CHARACTERS
%%%%%%%%%%%%%%%%%%%%%%%%%%%%%%%%%%%%%%%%%%%
\renewcommand{\z}{\int dz}
\def\h{\hbar}
\newcommand{\lp}{\left(}
\newcommand{\rp}{\right)}
\renewcommand{\lq}{\left[}
\renewcommand{\rq}{\right]}
\newcommand{\lgr}{\left\{}
\newcommand{\rgr}{\right\}}
\newcommand{\identity}{1\hspace{-0.4em}1}
\newcommand{\no}{\nonumber}
\newcommand{\ph}{\phantom} 
\def\abs#1{\left|#1\right|}
\def\tr{\,\mbox{Tr}\,}
\def\frac#1#2{ {{#1} \over {#2} }}
\def\fracd#1#2{ {\displaystyle {{#1} \over {#2} }}}
\def\ed#1{\displaystyle{e^{#1}}}
\def\half{\mbox{\small $\frac{1}{2}$}}
\def\p{\partial}
% \dpad {}{} is partial deriv over partial derivative
\newcommand{\dpad}[2]{{\displaystyle{\frac{\partial #1}{\partial #2}}}}
% \dfud {}{} is delta {} over delta {}
\newcommand{\dfud}[2]{{\displaystyle{\frac{\delta #1}{\delta #2}}}}
\def\ie{\hbox{\it i.e.}{ }}      
\def\etc{\hbox{\it etc.}{ }}
\def\eg{\hbox{\it e.g.}{ }}      
\def\bom#1{\mbox{\boldmath$#1$}}
\def\en{\bom{n}}
\def\gq{g^4}
\def\kap{\bom{k}}
\begin{frontmatter}
\title{Multiple vacua in two-dimensional Yang-Mills theory}
\author{A. Bassetto and F. Vian }
\address{Dipartimento di Fisica ``G. Galilei'' and INFN, Sezione di Padova,\\
via Marzolo 8, 35131 Padua, Italy}
\author{L. Griguolo}
\address{Dipartimento di Fisica ``M. Melloni'' and INFN, Gruppo Collegato 
di Parma,\\
viale delle Scienze, 43100 Parma, Italy}
\date{21 July 2000}
\begin{abstract}
Two-dimensional $SU(N)$ Yang-Mills theory is endowed with a non-trivial vacuum 
structure ($k$-sectors). The presence of $k$-sectors modifies
the energy spectrum of the theory and its instanton content, the
(Euclidean) space-time being  compactified on a sphere. For the
exact solution, in the limit in which the sphere is decompactified,
a $k$-sector can be mimicked by the presence of $k$-fundamental charges at
$\infty$, according to a Witten's suggestion. However, this property neither
holds before decompactification nor for the genuine perturbative solution 
which corresponds to the zero-instanton contribution on the sphere.  
\end{abstract}
\end{frontmatter}
\section{Introduction}
The vacuum structure in non-abelian quantum gauge theories is
still far from being satisfactorily understood.
Many important aspects of four-dimensional QCD, such as chiral symmetry 
breaking, are believed to be related to the existence of 
multiple vacuum states (labelled by a topological parameter, the
$\theta$ angle) and to quantum tunnelling between them mediated by
instanton effects. 
Unfortunately  four-dimensional QCD appears too 
complicated to deal with. 
Nevertheless, in the Light-Front 
(LF) formulation of a Quantum Field Theory, where the theory is
quantized not  on a space-like 
surface but on a light-like one (see \cite{light} for an extensive
review), at least in principle, 
the exact ground state is described as a simple Fock vacuum.
Actually,
in four-dimensional gauge theories such an approach is far from being 
simple, due to the intricate dynamics of the so-called
zero-modes. This  problem does not exist in two dimensions
(zero-modes are related to transverse degrees of freedom in LF quantization) 
and recently it has 
been shown \cite{capo} that, at the perturbative level, LF
quantization encodes a complicated instanton dynamics present in the 
equal-time (ET) formulation. 
Therefore two-dimensional gauge theories candidate  themselves as the 
simplest models in which the influence of 
topological parameters on physical quantities 
 can be probed.  
\section{\kap-sectors and instantons}
The best example of a theory which admits multiple vacua and shares 
relevant features with four-dimensional gauge theories is $QCD_2$
with  adjoint 
fermions, as noticed many years ago by Witten \cite{wittheta}.
What is remarkable is that adjoint 
matter in some sense mimics
transverse degrees of freedom, thus inducing a complex 
behaviour.
We limit ourselves  to investigate the $SU(N)$ case when fermionic 
dynamics is essentially frozen, considering infinitely massive adjoint quarks 
and studying the static potential between them.  
Since Yang-Mills fields
transform in the adjoint representation, the true local symmetry is
the quotient of $SU(N)$ by its center, $Z_N$. A standard result in
homotopy theory tells us that the quotient is not simply
connected, the first homotopy group being $\Pi_1(SU(N)/Z_N)=Z_N.$
This result is of particular relevance for the vacuum structure of a
two-dimensional gauge theory: in the case at hand we have
exactly  $N$ inequivalent quantizations, parametrized by 
a single integer $k$, taking the values $k=0,1,..,N-1$. 
Concerning the pure $SU(N)$ Yang-Mills theory, the  explicit solution 
when $k$-states are taken into account was presented in
Ref.~\cite{canad}:  their 
main result, the heat-kernel propagator on the cylinder, allows to compute 
partition functions and Wilson loops winding around a smooth non
self-intersecting closed contour on the sphere $S^2$.
Wilson loops, in this case,  
strongly depend on $k$
and, for the theory defined on the plane, read
\begin{eqnarray}
\label{wkdecfin}
&&\mathcal{W}_{k} (A)=\frac1{N^2-1} \left[ 1+
\frac{k N (N+2)(N-k)}{(k+1)(N-k+1)}
e^{-\frac{g^2 A}2 \,(N+1)}
\right. \no  \\
&+& \left.
\frac{(N+1)(N-k-1)}{k+1}
e^{-\frac{g^2 A}2 \,(N-k)}+ 
\frac{(N+1)(k-1)}{N-k+1}
e^{-\frac{g^2 A}2 \,k}\right]\,,
\end{eqnarray}
$A$ being the area singled out by the loop.
The result \re{wkdecfin} can be obtained starting from the true
$SU(N)/Z_N$  theory on the sphere \cite{bgv2}, in the
decompactification limit or  directly on the plane, using the
procedure of  \cite{paniak}, working with 
$SU(N)$ and simulating the $k$-sectors with a Wilson loop at infinity in the 
$k-$fundamental representation. 

On the other hand we checked that the very same result can be obtained 
through a perturbative resummation \cite{bgv2} 
of the expectation value of an adjoint loop enclosed in 
an asymptotic  $k$-fundamental loop  on the plane (at least up to
$O(g^4)$), when  the theory is quantized in the light-cone gauge
$A_{-}=0$ and
on the light front (LF). 
Hence the perturbative LF formulation seems to capture  the exact
result, even in presence of a non-trivial topology.
%
%\section{The instanton representation}
%

As first pointed out by Witten \cite{witte}, one can represent  the
Wilson loop $\mathcal{W}_k$ 
on $S^2$ 
as a sum over instanton contributions.
By instanton we mean a non-trivial classical solution of the
Yang-Mills  equations on $S^2$, which takes the form of an Abelian
Dirac  monopole embedded into the non-abelian gauge group.
At variance with the $SU(N)$ or $U(N)$ case, where any of such
configurations is characterized by a set of $N$ integers
$(n_1,\ldots,n_N)$,  in the $SU(N)/Z_N$ case,
by  Poisson-resumming  on the sphere, instanton numbers are seen to be 
generalized to rational values by the effect of $k$.
The set $(0,\ldots,0)$, representing the topologically trivial
solution, turns out to be  reproduced  if integration over the group  manifold
is replaced by integration over the tangent group algebra.
In Ref.~\cite{capo} it was shown,
for the case of a Wilson loop in the fundamental representation
of the group $U(N)$, that the zero-instanton sector can be obtained by
a {\it bona fide} perturbative calculation 
for the theory quantized in the light-cone
gauge by means of equal-time (ET) canonical commutators.
In the case of a Wilson loop 
in the adjoint representation for the $SU(N)$ theory, with $k$-sectors  
taken into account, an intriguing interplay occurs
between instantons and $k$-states  and we expect that the truly perturbative 
physics ignore the existence of the $k$ parameter.
In fact the zero-instanton limit does not depend on $k$ and still
reproduces  the ET
computation (in the decompactification limit) {\it without} the loop
at  infinity 
\cite{bgv2}
\begin{eqnarray}
\label{wilzero}
&&\mathcal{W}_{k}^{(0)}(A)=\frac{1}{N+1}+\frac{N}{\mathcal{Z}
\,(N+1)}\int_{-\infty}^{+\infty}dz_1\ldots dz_N
\exp \left[ -\frac{1}{2}\sum_{j=1}^{N}z_j^2 \right] \times\nonumber \\
&&\exp\lq ig(z_1-z_2)
\sqrt\frac{A}{2}\rq \Delta^2(z_1,\ldots,z_N),
\end{eqnarray}
where $\mathcal{Z}=\int \mathcal{D}F\,
\exp(-\frac{1}{2}{\rm Tr}F^2).$
%
%\section{Two-loop correlation on the sphere}
%

Next, we examined  the correlation on the sphere
between two non-intersecting (nested) loops, one in the adjoint representation
and the other in the $k$-fundamental one. 
Our purpose in so doing was to explore to what extent this procedure
reproduces the result we  obtained working with a
single loop in the adjoint representation on the sphere in a $k$-sector.
We found
that Witten's conjecture, namely, in our language, that 
the $SU(N)/Z_N$ theory on the plane in a $k$-sector is equivalent to the 
usual $SU(N)$ theory in presence of a $k$-fundamental Wilson loop at 
infinity, is indeed
verified. 
However this is true only for the exact solution,
and only after decompactification of the sphere.

We now address the issue of singling the zero-instanton (trivial) sector out
for the adjoint loop enclosed in a $k$-fundamental one. 
The result we obtained  \cite{bgv2} shows the following basic features:
\begin{enumerate}
\item
for $k \neq 0$, it is different from the single loop case;
\item 
although string tensions are independent of $k$, 
the polynomial coefficients do  depend on it.
\end{enumerate}
One should not be too surprised that
the instanton structures of the two cases, as long as one remains 
on the sphere, are completely different. 
${\it Only}$ for the complete theory on the 
plane (i.e. full-instanton resummed and then decompactified) the equivalence 
between $k$-sectors and theories with $k$-fundamental Wilson loops at
infinity  holds; but
 there is no reason why this miracle should occur
when the two (different) zero-instanton sectors are compared.

We finally checked that the zero-instanton contribution to 
the expectation value of an adjoint loop enclosed in
a  $k$-fundamental one  on the plane is consistent, at least up to
$O(g^4)$,  with the
perturbative computation where the ET propagator is used.

At this stage, we think we have set a solid ground for the most
interesting  future development, namely the introduction of dynamical
fermions, with a particular focus on the generation and on the
properties of  a chiral condensate.
\end{document}